\documentclass[12pt]{amsart}
\usepackage{amsthm}
\advance\textheight\topskip
\usepackage{times}

\usepackage[mathscr]{eucal}
\usepackage{graphicx}

\usepackage{pb-diagram}

\usepackage[T1]{fontenc}

\usepackage{hyperref}

\allowdisplaybreaks

\newcommand{\Tr}{\mathop{\mathrm{Tr}}\nolimits}

\newcommand{\ffrac}[2]{\raisebox{.6pt}{\mbox{%
    \footnotesize$\displaystyle\frac{#1\mathstrut}{#2\mathstrut}$}}}

\newcommand{\fhalf}{\ffrac{1}{2}}

\newcommand{\mffpp}{\mathrm{MFF}^{+}_{\!(+)}}
\newcommand{\mffpm}{\mathrm{MFF}^{+}_{\!(-)}}
\newcommand{\mffmp}{\mathrm{MFF}^{-}_{\!(+)}}
\newcommand{\mffmm}{\mathrm{MFF}^{-}_{\!(-)}}
\newcommand{\mffpPM}{\mathrm{MFF}^{+}_{\!(\pm)}}

\newcommand{\mffold}[1]{\mathrm{MFF}^{#1}}

\def\half{\frac{1}{2}}

\newcommand{\Cminus}{C^{(-)}}
\newcommand{\cCminus}{\mathcal{C}^{(-)}}
\newcommand{\Cplus}{C^{(+)}}
\newcommand{\cCplus}{\mathcal{C}^{(+)}}

\numberwithin{equation}{section}
\makeatletter
\@addtoreset{equation}{section}

\makeatother

\newtheorem{Thm}{Theorem}[section]
\newtheorem{Lemma}[Thm]{Lemma}

\theoremstyle{definition}
\newtheorem{Dfn}[Thm]{Definition}
\newtheorem{Rem}[Thm]{Remark}%[section]

\newcommand{\hSSL}[2]{\widehat{s\ell}(#1|#2)}
\newcommand{\SSL}[2]{s\ell(#1|#2)}
\newcommand{\hSL}[1]{\widehat{s\ell}(#1)}

\newcommand{\N}[1]{N\!=\!#1}

\newcommand{\module}{\mathscr}
\def\mM{\module{M}}
\newcommand{\mP}{\module{P}}
\newcommand{\mR}{\module{R}}

\newcommand{\verma}[1]{\module{P}_{#1}}
\newcommand{\vermaNplus}[1]{\module{N}_{#1}^+}
\newcommand{\vermaNminus}[1]{\module{N}_{#1}^-}
\newcommand{\vermaNpm}[1]{\module{N}_{#1}^{\pm}}
\newcommand{\vermaNmp}[1]{\module{N}_{#1}^{\mp}}

\newcommand{\ket}[1]{|{#1}\rangle}
\newcommand{\ketplus}[1]{|{#1}\rangle^+}
\newcommand{\ketminus}[1]{|{#1}\rangle^-}

\newcommand{\Hminus}{H^-}
\newcommand{\Hplus}{H^+}

\newcommand{\oC}{\mathbb{C}}
\newcommand{\oN}{\mathbb{N}}

\newcommand{\oZ}{\mathbb{Z}}

\newcommand{\jplus}{\mathsf{j}^+}
\newcommand{\jminus}{\mathsf{j}^-}

\def\cA{\mathcal{A}}

\def\cC{\mathcal{C}}

\def\cL{\mathcal{L}}

\def\cU{\mathcal{U}}

\def\cX{\mathcal{X}}

\def\rom#1{{\rm #1}}

\begin{document}
\raggedbottom

\title{Twists and singular vectors in $\widehat{s\ell}(2|1)$
  representations}

\author[Semikhatov]{A.~M.~Semikhatov}
\address{Lebedev Physics Institute, Russian Academy of
  Sciences, Moscow, Russia}
\author[Taormina]{A.~Taormina}
\address{Department of Mathematical Sciences, University of Durham,
  UK}

\begin{abstract}
  We propose new formulas for singular vectors in Verma modules over
  the affine Lie superalgebra $\hSSL21$.  We analyze the coexistence
  of singular vectors of different types and identify the twisted
  modules $\vermaNpm{h,k;\theta}$ arising as submodules and quotient
  modules of $\hSSL21$ Verma modules. We show that with the twists
  (spectral flow transformations) properly taken into account, a
  resolution of irreducible representations can be constructed
  consisting of only the $\vermaNpm{h,k;\theta}$ modules.
\end{abstract}

\maketitle

\thispagestyle{empty}

\section{Introduction}
In this paper, we consider elements of representation theory of the
affine Lie superalgebra $\hSSL21$.  The affine $\SSL21$ symmetry
emerges in the models of disordered systems introduced in relation to
the integer quantum Hall effect~\cite{1}--\cite{3}.  The $\hSSL21$
algebra is also interesting because in a certain sense, it combines
the characteristic features of the affine Lie algebra $\hSL2$ and the
$\N2$ superconformal algebra.  To the latter, it is related via
Hamiltonian reduction~\cite{4}--\cite{6} and its
``inversion''~\cite{7}; a relation between the $\hSSL21$ and $\hSL2$
algebras, apart from the obvious subalgebra embedding, was worked out
in~\cite{8}, where $\hSSL21$ was shown to be a vertex-operator
extension of the sum of two $\hSL2$ algebras with ``dual'' levels $k$
and $k'$ such that $(k+1)(k'+1)=1$.  The construction in~\cite{8} has
led to a decomposition formula for $\hSSL21$ representations in terms
of $\hSL2_{k}\oplus\hSL2_{k'}$ representations.  A natural class of
$\hSSL21$ representations can be obtained by taking admissible
representations of the two $\hSL2$ algebras.  However, by far not all
of these $\hSSL21$ representations have been studied before; in
particular, the corresponding characters are only known for a subclass
of these representations~\cite{9}.

Characters of a broader class of $\hSSL21$ representations can be
found by constructing resolutions; this in turn requires analyzing
singular vectors in and the mappings between Verma modules.  An
important role is here played by spectral flows (\textit{twists}) and
other automorphisms of the $\hSSL21$ algebra.  In this paper, we
generalize the ``continued'' formulas for singular
vectors~\cite{9},~\cite{10} in $\hSSL21$ Verma modules such that the
new formulas are applicable to the case where degenerations of a
certain type occur, under which the previously known formulas gave the
(incorrect) vanishing result.  These are the degenerations where the
so-called ``charged'' singular vector exists simultaneously with the
MFF singular vectors to be introduced below.  This ``stability'' of
singular vectors under degenerations of modules is ensured by
incorporating twists into the ``continued'' formula.  These singular
vectors can be used in constructing the resolutions and finding
character formulas.

\begin{figure}[tb]
  \begin{center}
    \includegraphics[angle=270]{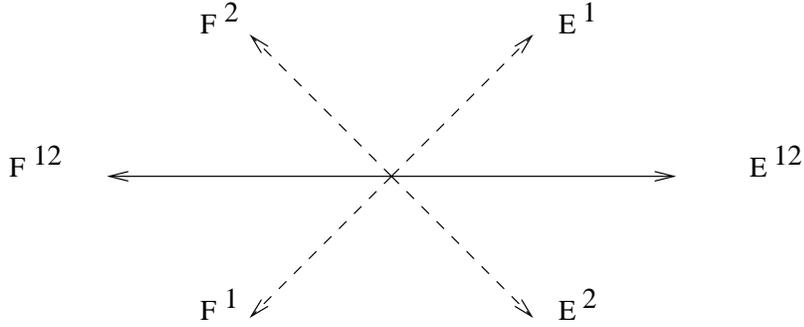}
  \end{center}
  \caption[Root diagram of $s\ell(2|1)$.]{\textsl{Root diagram of
      $s\ell(2|1)$.}}
  \label{fig:sl21-roots}
\end{figure}

\section{$\hSSL21$ modules and automorphisms} \label{sec:auto}
\subsection{The $\protect\widehat{s\ell}(2|1)$ algebra and
  automorphisms}\label{app:sl21} The affine Lie superalgebra $\hSSL21$
is spanned by four bosonic currents $E^{12}$, $H^-$, $F^{12}$, and
$H^+$, four fermionic ones, $E^1$, $E^2$, $F^1$, and $F^2$, and the
central element (which we identify with its eigenvalue~$k$).  For
convenience, we give in Fig.~\ref{fig:sl21-roots} the two-dimensional
root diagram of the finite-dimensional Lie superalgebra $s\ell(2|1)$,
represented in Minkowski space with the fermionic roots along the
light-cone directions.  The $\hSL2$ subalgebra is generated by
$E^{12}$, $H^-$, and $F^{12}$, and it commutes with the~$u(1)$
subalgebra generated by $H^+$.  The nonvanishing commutation relations
are given by
%\begin{equation}
\begin{alignat*}{2}
  &{[}H^-_m, E^{12}_n] = E^{12}_{m+n},&\qquad &{[}H^-_m, F^{12}_n] =
  -F^{12}_{m+n},
  \\
  &{[}E^{12}_m, F^{12}_n] = m \delta_{m+n, 0} k + 2 H^-_{m+n},&\qquad
  &{[}H^\pm_m, H^\pm_n] = \mp\fhalf m \delta_{m+n, 0} k,
  \\
  &{[}F^{12}_m, E^2_n] = F^1_{m+n},&\qquad &{[}E^{12}_m, F^2_n] =
  -E^1_{m+n},
  \\
  &{[}F^{12}_m, E^1_n] = -F^2_{m+n},&\qquad &{[}E^{12}_m, F^1_n] =
  E^2_{m+n},
  \\
  &{[}H^\pm_m, E^1_n] = \fhalf E^1_{m+n},&\qquad &{[}H^\pm_m, F^1_n] =
  -\fhalf F^1_{m+n}, \label{2.1}
  %\label{sl21}
  \\
  &{[}H^\pm_m, E^2_n] = \mp\fhalf E^2_{m+n},&\qquad &{[}H^\pm_m, F^2_n]
  = \pm\fhalf F^2_{m+n},
  \\
  &{[}E^1_m, F^1_n] = -m \delta_{m+n, 0} k + H^+_{m+n} -H^-_{m+n},&&
  \\
  &{[}E^2_m, F^2_n] = m \delta_{m+n, 0} k + H^+_{m+n} +H^-_{m+n},&&
  \\
  &{[}E^1_m, E^2_n] = E^{12}_{m+n},&\qquad &{[}F^1_m, F^2_n] =
  F^{12}_{m+n}.
\end{alignat*}
%\end{equation}
The Sugawara energy-momentum tensor is given by
\begin{equation}%\label{Tsug-sl21}
  T_{\mathrm{Sug}} =
  \ffrac{1}{k +1}\bigl(\Hminus \Hminus - \Hplus \Hplus +
  E^{12} F^{12} + E^1 F^1 - E^2 F^2 \bigr).
  \label{2.2}
\end{equation}

There are the algebra automorphisms
\begin{align}
  \alpha&:
  \begin{aligned}
    E^1_n\mapsto{}& F^2_n,&\quad E^2_n\mapsto{}& F^1_n,&\quad
    E^{12}_n\mapsto{}& F^{12}_n,
    \\
    F^1_n\mapsto{}& E^2_n,& F^2_n\mapsto{}& E^1_n,&
    F^{12}_n\mapsto{}& E^{12}_n,
    \\
    H^+_n\mapsto{}& H^+_n, &H^-_n\mapsto{}&{-H^-_n},
  \end{aligned}
  \label{2.3}
  \\[6pt]
  \beta&:
  \begin{aligned}
    E^1_n\mapsto{}& E^2_n,&\quad E^2_n\mapsto{}& E^1_n,
    &E^{12}_n\mapsto{}& E^{12}_n,
    \\
    F^1_n\mapsto{}& -F^2_n,& F^2_n\mapsto{}& -F^1_n,&\quad
    F^{12}_n\mapsto{}& F^{12}_n,
    \\
    H^+_n\mapsto{}& -H^+_n,& H^-_n\mapsto{}& H^-_n,
  \end{aligned} %\label{beta}
  \label{2.4}
\end{align}
and a family of automorphisms for $\theta\in\oZ$,
\begin{equation}
  \cU_\theta:
  \begin{aligned}
    E^1_n\mapsto{}& E^1_{n-\theta},& \quad E^2_n\mapsto{}&
    E^2_{n+\theta},
    \\
    F^1_n\mapsto{}& F^1_{n+\theta}, &F^2_n\mapsto{}& F^2_{n-\theta},
  \end{aligned}
  \quad H^+_n\mapsto H^+_n + k\theta\delta_{n,0}
  %\label{sl21-spectral}
  \label{2.5}
\end{equation}
(with the $\hSL2$ subalgebra remaining invariant).  These
automorphisms satisfy the relations
\begin{equation}
  \alpha^2=1,\qquad\beta^2=1,\qquad(\alpha\beta)^4=1,\qquad
  \alpha\cU_\theta=\cU_\theta\alpha,\qquad(\beta\cU_\theta)^2=1.
  \label{2.6}
\end{equation}
The $\oZ$ subgroup of automorphisms $(\cU_\theta)_{\theta\in\oZ}$ is
called the \textit{spectral flow} and is extensively used in what
follows.  Another $\oZ$ algebra of automorphisms (a spectral flow
affecting the $\hSL2$ subalgebra, cf.~\cite{11}) acts as
\begin{equation}%\label{sl2-spectral}
  \cA_\eta:{}
  \begin{aligned}
    E^1_n\mapsto{}& E^1_{n+\eta},& E^2_n\mapsto{}&
    E^2_{n+\eta},&~ E^{12}_n\mapsto{}& E^{12}_{n+2\eta},&~
    H^-_n\mapsto{}& H^-_n + k\eta\delta_{n,0},
    \\
    F^1_n\mapsto{}& F^1_{n-\eta}, &F^2_n\mapsto{}&
    F^2_{n-\eta}, &F^{12}_n\mapsto{}& F^{12}_{n-2\eta},
    &H^+_n\mapsto{}& H^+_n.
  \end{aligned}
  \label{2.7}
\end{equation}

There also is an \textit{automorphism}
\begin{equation}
  \gamma=\cU_{\half}\circ\cA_{-\half}
  \label{2.8}
\end{equation}
(while $\cU_{\half}$ and $\cA_{-\half}$ are not automorphisms, but
mappings into an isomorphic algebra, their composition is).  For
$\theta\in\oZ$, its powers $\cX_\theta=\gamma^\theta$ map the
generators~as
\begin{gather}%\label{exotic-spectral}
  \cX_\theta:{}
  \begin{aligned}
    E^1_n\mapsto{}& E^1_{n-\theta},&E^2_n\mapsto{}&E^2_{n},&~
    E^{12}_n\mapsto{}& E^{12}_{n-\theta},&~ H^-_n\mapsto{}& H^-_n -
    \ffrac{k}{2}\theta\delta_{n,0},
    \\
    F^1_n\mapsto{}& F^1_{n+\theta}, &F^2_n\mapsto{}& F^2_{n},
    &F^{12}_n\mapsto{}& F^{12}_{n+\theta}, &H^+_n\mapsto{}& H^+_n +
    \ffrac{k}{2}\theta\delta_{n,0}.
  \end{aligned}
  \label{2.9}
\end{gather}

\subsection{Twisted highest-weight conditions and twisted Verma
  modules} Applying algebra automorphisms to modules gives generically
non-isomorphic modules.  The nilpotent subalgebra is also mapped under
the action of automorphisms, and the annihilation conditions satisfied
by highest-weight vectors change accordingly.  Thus, the existence of
an automorphism group entails a certain freedom in choosing the type
of annihilation conditions imposed on highest-weight vectors in
highest-weight (in particular, Verma) modules.  We select a family of
nilpotent subalgebras such that the corresponding annihilation
conditions read\footnote{The $\approx$ sign means that the operators
  must be applied to a vector; at the moment, we are interested in the
  list of annihilation operators, rather than in the vector, hence the
  notation.}
\begin{equation}%\label{twisted-hwcond}
  \begin{aligned}
    E^1_{\geq-\theta}\approx{}& 0,& E^2_{\geq\theta}\approx{}&0,
    \\
    F^1_{\geq\theta+1}\approx{}&0,& F^2_{\geq1-\theta}\approx{}&0,
    \qquad F^{12}_{\geq1}\approx0
  \end{aligned}
  \label{2.10}
\end{equation}
for a fixed $\theta\in\oZ$.  These annihilation conditions are called
the \textit{twisted highest-weight conditions} for~$\theta\neq0$ and
the untwisted ones in the particular case where~$\theta=0$.  The
twisted highest-weight conditions with different $\theta$ are mapped
into one another by spectral flow~\eqref{2.5}.

We note that the automorphism $\gamma\circ\alpha$ maps the twisted
highest-weight conditions~\eqref{2.10} into highest-weight conditions
of a different class, namely those where the annihilation conditions
are given~by
\begin{gather}%\label{second-hwcond}
  \begin{aligned}
    E^1_{\geq-\theta}\approx{}&0,\quad & E^2_{\geq\theta+1}\approx&{}
    0,\qquad E^{12}_{\geq0}\approx0,
    \\
    F^1_{\geq\theta+1}\approx{}&0,&F^2_{\geq-\theta}\approx{}&0.
  \end{aligned}
  \label{2.11}
\end{gather}
A Verma module generated from a twisted highest-weight state
(satisfying conditions~\eqref{2.10}) can contain a \textit{submodule}
generated from a state satisfying~\eqref{2.11}, and it is not a
priori guaranteed that the same submodule can be generated from a
twisted highest-weight state satisfying annihilation
conditions~\eqref{2.10} for some~$\theta$.  This is in contrast with
the more familiar case of the $\hSL2$ algebra, where any submodule in
a given Verma module can be generated from vector(s) satisfying
highest-weight conditions of the same type as for the highest-weight
vector of the module (i.e., \textit{not} those transformed by
automorphisms).  Similarly, even within the chosen
family~\eqref{2.10} of twisted highest-weight conditions, a module
with the untwisted ($\theta=0$) highest-weight vector can have
submodules generated from some twisted highest-weight states, but
\textit{not} from the one with zero twist.

For any $\hSSL21$ module $\mR$, we let $\mR_{;\theta}$ denote the
twisted module $\cU_{\theta}\mR$, see~\eqref{2.5}.

\subsection{The action of the spectral flow transform on characters}
Spectral flow transform~\eqref{2.5} acts on characters as follows.
The character
$\chi^{\phantom{y}}_{\mR_{;\theta}}\equiv\chi^{\mR}_{;\theta}$ of a
twisted module ${\mR}_{;\theta}$ is expressed through the character of
$\mR$ as
\begin{equation}%\label{sl21-sf}
  \chi^{\mR}_{;\theta}(q,z,\zeta)=
  \zeta^{-k\theta}\,q^{-k\theta^2}\,\chi^{\mR}(q,z,\zeta\,q^{2\theta}).
  \label{2.12}
\end{equation}

Clearly, in studying characters (as well as other properties of
representations), it therefore suffices to consider a representative
of each spectral flow orbit.

\subsection{Gradings and extremal states} 
\begin{figure}[bh]
  \begin{center}
    \includegraphics[trim=1cm 1cm -9cm -9cm, clip, scale=.5]{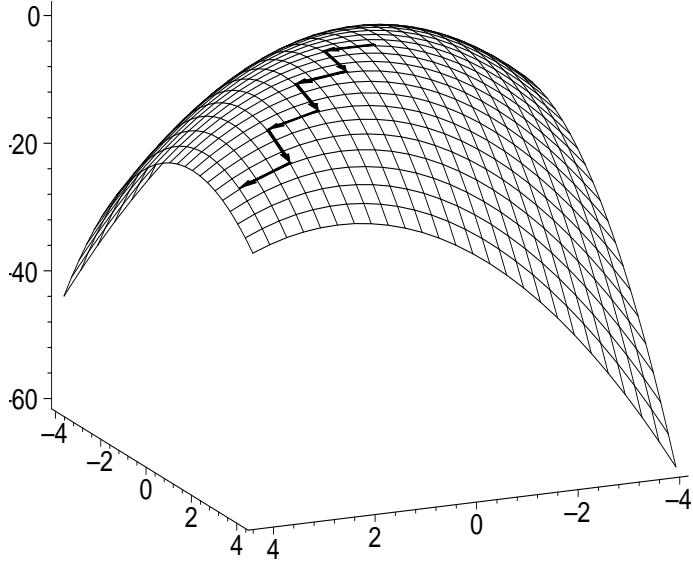}
  \end{center}
  \caption[Extremal diagram of an  $\hSSL21$ Verma
  module.]{\textsl{Extremal diagram of an $\hSSL21$ Verma module}.
    \small The paraboloid surface interpolates between the extremal
    states.  The horizontal axes represent the eigenvalues of $H^+_0 +
    H^-_0$ and $H^+_0 - H^-_0$ and the vertical axis gives minus the
    level; the highest-weight vector is conventionally placed at the
    origin.  The sequence of arrows shows a charged singular vector
    (see\ Sec.~\ref{sec:charged}); depending on how the
    identifications are made, this is either $\Cminus(-3)$ (with each
    odd arrow representing an $F^1_n$ and each even arrow an~$E^2_m$
    generator) or $\Cplus(3)$ (respectively, $F^2$ and $E^1$), see
    Eqs.~\eqref{4.2} and~\eqref{4.3}.  The action of spectral
    flow~\eqref{2.5} results in moving the surface over itself such
    that the origin slides down along the parabolic curve running
    through the endpoints of every second arrow.}
  \label{fig:extremal}
\end{figure}

Clearly, each $\hSSL21$
module is $\oZ\times\oZ\times\oZ$-graded: the gradings are by the
\textit{charge} (the $H^-_0$ eigenvalue), the \textit{hypercharge}
(the $H^+_0$ eigenvalue), and the \textit{level} (minus the $L_0$
eigenvalue, where $L_n$ are the Virasoro generators corresponding to
the Sugawara energy-momentum tensor).  Thus, states of the module
occupy sites of a three-dimensional lattice.  Because of the
highest-weight conditions, all the states in the module lie below a
certain plane in the three-dimensional space.  Those lattice sites
that are occupied by at least one state from the module form a
(convex) three-dimensional body in $\oZ\times\oZ\times\oZ$.  The
states at the surface of this body are called \textit{extremal
  states}, or extremal vectors.  In particular, the highest-weight
state is an extremal vector.  In Fig.~\ref{fig:extremal}, we give the
extremal diagram of an $\hSSL21$ Verma module.

\section{$\hSSL21$ Verma Modules}
In what follows, we assume that $k\in\oC\setminus\{-1\}$, \ $h_-$ and
$h_+$ are a priori arbitrary complex numbers, and $\theta\in\oZ$.

We define a family of \textit{twisted} Verma modules related to each
other by the spectral flow.
\begin{Dfn}\rm
  A twisted Verma module $\verma{h_-,h_+,k;\theta}$ over the level-$k$
  $\hSSL21$ algebra is freely generated by $E^1_{\leq-\theta-1}$,
  $E^2_{\leq\theta-1}$, $E^{12}_{\leq-1}$, $F^1_{\leq\theta}$,
  $F^2_{\leq-\theta}$, $F^{12}_{\leq0}$, $H^-_{\leq-1}$, and
  $H^+_{\leq-1}$ from the twisted highest-weight state
  $\ket{h_-,h_+,k;\theta}$ satisfying annihilation
  conditions~\eqref{2.10} and the conditions
  \begin{align}%\label{twisted-eigen-minus}
    &H^-_0\,\ket{h_-,h_+,k;\theta} = {}h_-\,\ket{h_-,h_+,k;\theta},
    \label{3.1}
    \\
    &H^+_0\,\ket{h_-,h_+,k;\theta} = (h_+ -
    k\theta)\,\ket{h_-,h_+,k;\theta}.
    %\label{twisted-eigen-plus}
    \label{3.2}
  \end{align}
\end{Dfn}

We then have
\begin{equation}
  \cU_{\theta'}\ket{h_-,h_+,k;\theta}=\ket{h_-,h_+,k;\theta+\theta'}
  \label{3.3}
\end{equation}
and, obviously, $\cU_{\theta'}\verma{h_-,h_+,k;\theta}=
\verma{h_-,h_+,k;\theta+\theta'}$ for Verma modules.

We write
\begin{equation}%\label{doteq}
  \ket{X}\doteq\ket{h_-,h_+,k;\theta}
  \label{3.4}
\end{equation}
for any state $\ket{X}$ that satisfies highest-weight
conditions~\eqref{2.10} and \eqref{3.1}--\eqref{3.2}.

The character of $\verma{h_-,h_+,k;\theta}$ is given by
\begin{multline}%\label{sl21-Verma-char}
  \label{3.5}
  \chi^{\verma{}}_{h_-,h_+,k;\theta}(q,z,\zeta)\equiv
  \Tr^{\phantom{y}}_{\verma{h_-,h_+,k;\theta}}
  \Bigl(q^{\cL^{\text{Sug}}_0}\,z^{H^-_0}\,\zeta^{H^+_0}\Bigr)=
  \\
  = z^{h_-}\,\zeta^{h_+-(k+1)\theta}\, q^{\frac{h_-^2 - h_+^2}{k+1} +
    2\theta h_+ -(k+1)\theta^2}\, \frac{\vartheta_{1,0}(q,
    z^{\half}\zeta^{\half})\, \vartheta_{1,0}(q,
    z^{\half}\zeta^{-\half})}
  {\vartheta_{1,1}(q,z)\prod_{m\geq1}(1-q^m)^3},  
\end{multline}
where the Jacobi theta functions are defined by
\begin{multline}%\label{Jacobi11}
  \label{3.6}
  \vartheta_{1,1}(q, z) = \sum_{m\in\oZ}(-1)^m q^{\half(m^2 -m)}
  z^{-m} \\*
  \smash[b]{{}=\prod_{m\geq0}(1 - z^{-1} q^m) \prod_{m\geq1}(1 - z
    q^m)\prod_{m\geq1}(1 - q^m)},
\end{multline}
\begin{multline}
%\label{Jacobi10}
  \vartheta_{1,0}(q,z) = \sum_{m\in\oZ}^{}q^{\half(m^2 - m)} z^{-m}\\*
  =
  \prod_{m\geq0}(1+z^{-1}q^m)\prod_{m\geq1}(1+z q^m)
  \prod_{m\geq1}(1-q^m).  \label{3.7}
\end{multline}

We now let $\ketminus{h_-,k;\theta}$ denote the state satisfying the
highest-weight conditions
\begin{equation}%\label{hwtop-minus}
  \begin{aligned}
    E^1_{-\theta}\approx{}&0,& \quad E^2_{\theta}\approx{}&0,
    \\
    F^1_{\theta}\approx{}&0,&\quad F^2_{1-\theta}\approx{}&0
  \end{aligned}
  \label{3.8}
\end{equation}
and
\begin{equation}%\label{hwtop-eigen}
  H^-_0\ketminus{h_-,k;\theta}= h_-\ketminus{h_-,k;\theta}.
  \label{3.9}
\end{equation}
Comparing the annihilation conditions in Eqs.~\eqref{2.10}
and~\eqref{3.8}, we see that the latter are a strengthened version of
the former.  In view of the commutation relations of the algebra, it
then follows that only one of the two highest-weight parameters
$h_\pm$ remains independent.  We let $\vermaNminus{h_-,k;\theta}$ be
the module freely generated by $E^1_{\leq-\theta-1}$,
$E^2_{\leq\theta-1}$, $E^{12}_{\leq-1}$, $F^1_{\leq\theta-1}$,
$F^2_{\leq-\theta}$, $F^{12}_{\leq0}$, $H^-_{\leq-1}$, and
$H^+_{\leq-1}$ from $\ketminus{h_-,k;\theta}$.  Its character is given
by
\begin{equation}%\label{char-minus}
  \chi^{\vermaNminus{}}_{h_-,k;\theta}(q,z,\zeta)=
  \frac{\chi^{\verma{}}_{h_-,h_-,k;\theta}(q,z,\zeta)}
  {1 + q^{-\theta} z^{-\half}\zeta^{-\half}}.
  \label{3.10}
\end{equation}

In terms of extremal diagrams of the type shown in Fig.~2, the module
$\vermaNminus{h_-,k;\theta}$ is more narrow than the Verma module
$\verma{}$.  Accordingly, we call $\vermaNminus{h_-,k;\theta}$ the
\textit{narrow} Verma modules (even though the term ``Verma module''
involves a certain abuse in this case).

There is the second type of narrow Verma modules.  Namely, let
$\ketplus{h_-,k;\theta}$ denote the state satisfying the
highest-weight conditions
\begin{equation}%\label{hwtop-plus}
  \begin{aligned}
    E^1_{-\theta}\approx{}&0,& \quad E^2_{\theta}\approx{}&0,
    \\
    F^1_{\theta+1}\approx{}&0,& F^2_{-\theta}\approx{}&0
  \end{aligned}
  \label{3.11}
\end{equation}
and
\begin{equation}
  H^-_0\ketplus{h_-,k;\theta}= h_-\ketplus{h_-,k;\theta},
  \label{3.12}
\end{equation}
and let $\vermaNplus{h_-,k;\theta}$ be the module freely generated by
$E^1_{\leq-\theta-1}$, $E^2_{\leq\theta-1}$, $E^{12}_{\leq-1}$,
$F^1_{\leq\theta}$, $F^2_{\leq-\theta-1}$, $F^{12}_{\leq0}$,
$H^-_{\leq-1}$, and $H^+_{\leq-1}$ from $\ketplus{h_-,k;\theta}$.  The
character of $\vermaNplus{h_-,k;\theta}$ is given by
\begin{equation}%\label{char-plus}
  \chi^{\vermaNplus{}}_{h_-,k;\theta}(q,z,\zeta)=
  \frac{\chi^{\mP{}}_{h_-,-h_-,k;\theta}(q,z,\zeta)}
  {1 + q^{\theta} z^{-\half}\zeta^{\half}}.
  \label{3.13}
\end{equation}

The states introduced above behave as follows under the action of some
of the automorphisms listed in Sec.~\ref{app:sl21}:
\begin{equation}
  \beta\ketplus{h_-,k;\theta}=\ketminus{h_-,k;-\theta},\qquad
  \cU_{\theta'}\ket{h_-,k;\theta}^\pm=\ket{h_-,k;\theta+\theta'}^\pm.
  \label{3.14}
\end{equation}
For the corresponding Verma modules, we then have, e.g.,
\begin{equation}%\label{beta-acts}
  \beta\,\vermaNpm{h_-,k;\theta}=\vermaNmp{h_-,k;-\theta}.
  \label{3.15}
\end{equation}
We also note that for any module $\mM$, the action of the automorphism
$\beta$ on its character is given by
$\chi^{\beta\,\mM}(q,z,\zeta)=\chi^{\mM}(q,z,\zeta^{-1})$.

\section{Charged singular vectors} \label{sec:charged}
The narrow Verma modules can be obtained from the Verma modules
$\verma{}$ by taking the quotient with respect to states that are
traditionally called the charged singular vectors.

\subsection{Charged singular vectors: explicit formulas}
By definition, \textit{charged singular vectors} are those singular
vectors that occur in the Verma module $\verma{h_-,h_+,k;\theta}$
whenever
\begin{equation}%\label{charged-exist}
  h_+ = \pm h_- - (k+1)n,\qquad n\in\oZ.
  \label{4.1}
\end{equation}
They are given by an explicit construction as follows~\cite{8},
\cite{10}.  For $h_+ -h_- = -(k+1)n$, $n\in\oZ$, the charged singular
vector in the twisted Verma module $\mP_{h_-,h_+,k;\theta}$ is given
by
\begin{multline}%\label{Echminus}
  \ket{\Cminus(n, h_-, k;\theta)}={}\\
  = \begin{cases}
      \underbrace{E^2_{\theta+n}\dots E^2_{\theta-1}}_{-n}\cdot
      \underbrace{F^1_{\theta+n}\dots F^1_{\theta}}_{-n+1}
       \ket{h_-,h_- - n(k+1),k;\theta}, &n\leq0,
      \\
      \underbrace{E^1_{-\theta-n}\dots E^1_{-\theta-1}}_{n}\cdot
      \underbrace{F^2_{1-\theta-n}\dots F^2_{-\theta}}_{n}
      \ket{h_-,h_- - n(k+1),k;\theta},& n\geq1.
    \end{cases}\kern-8pt
  \label{4.2}
\end{multline}
Similarly, whenever $h_+ +h_- = -n(k+1)$, $n\in\oZ$, the charged
singular vector in~$\mP_{h_-,h_+,k;\theta}$ is
\begin{multline}%\label{Echplus}
  \ket{\Cplus(n, h_-, k;\theta)} ={}\\
  = \begin{cases}
      \underbrace{E^2_{\theta+n}\dots E^2_{\theta-1}}_{-n}\cdot
      \underbrace{F^1_{\theta+n+1}\dots F^1_{\theta}}_{-n}
      \ket{h_-,-h_- - n(k+1),k;\theta},& n\leq-1,
      \\
      \underbrace{E^1_{-\theta-n}\dots E^1_{-\theta-1}}_{n} \cdot
      \underbrace{F^2_{-\theta-n}\dots F^2_{-\theta}}_{n+1}
      \ket{h_-,-h_- - n(k+1),k;\theta},& n\geq0.
    \end{cases}\kern-8pt
  \label{4.3}
\end{multline}

It is elementary to verify that vector~\eqref{4.2} satisfies the
annihilation conditions (we omit the singular vector itself)
\begin{equation}
  \begin{aligned}
    E^1_{-\theta-n}\approx{}&0,\quad& E^2_{\theta+n}\approx{}&0,
    \\
    F^1_{\theta+n}\approx{}&0,& F^2_{1-\theta-n}\approx{}&0,
  \end{aligned}
  \label{4.4}
\end{equation}
i.e., conditions~\eqref{3.8} with the twist parameter $\theta+n$.  It
is easy to see that these annihilation conditions imply that this
vector does indeed generate a submodule in the corresponding
module~$\mP_{h_-,h_+,k;\theta}$.  More precisely, we have
\begin{equation}
  \ket{\Cminus(n, h_-, k;\theta)}\doteq
  \begin{cases}
    \ketminus{h_- - \fhalf,k;\theta+n},& n\leq 0,
    \\
    \ketminus{h_-,k;\theta+n}, & n\geq1.
  \end{cases}
  \label{4.5}
\end{equation}
We here use the notation similar to~\eqref{3.4} to express the fact
that the vector in the left-hand side satisfies the same annihilation
conditions and has the same eigenvalues as those determined by
Eqs.~\eqref{3.8},~\eqref{3.9} for the vector in the right-hand side.

Similarly, vector~\eqref{4.3} satisfies the annihilation conditions
\begin{equation}
  \begin{aligned}
    E^1_{-\theta-n}\approx{}&0, &E^2_{\theta+n}\approx{}&0,
    \\
    F^1_{\theta+n+1}\approx{}&0,&\quad F^2_{-\theta-n}\approx{}&0,
  \end{aligned}
  \label{4.6}
\end{equation}
i.e., conditions~\eqref{3.11} with the twist parameter $\theta+n$, or
more precisely,
\begin{equation}
  \ket{\Cplus(n, h_-, k;\theta)}\doteq
  \begin{cases}
    \ketplus{h_-,k;\theta+n},& n\leq-1,
    \\
    \ketplus{h_- - \fhalf,k;\theta+n}, & n\geq0.
  \end{cases}
  \label{4.7}
\end{equation}

In what follows, we write 
\begin{gather*}
  C^{(\pm)}(n,h_-,k)=\cC^{(\pm)}(n,h_-,k)\,
  \ket{h_-,\mp h_- - n(k+1),k}
\end{gather*}
and call $\cC^{(\pm)}$ the \textit{charged singular vector operator}.
We now consider the submodules generated from $\ket{\Cminus(0, h_-,
  k;\theta)}$ and $\ket{\Cplus(0, h_-, k;\theta)}$ in more detail.

\subsection{Charged singular vectors: submodules and quotients} 
In the Verma module $\verma{h_-,h_-,k;\theta}$, there is the charged
singular vector $\ket{\Cminus(0, h_-, k;\theta)}$ given by
\begin{equation}%\label{F1}
  F^1_\theta\ket{h_-,h_-,k;\theta}\doteq
  \ketminus{h_--\fhalf,k;\theta}.
  \label{4.8}
\end{equation}
The submodule generated from~\eqref{4.8} is
$\vermaNminus{h_--\half,k;\theta}\subset\verma{h_-,h_-,k;\theta}$.
Moreover, we have the exact sequence
\begin{equation}%\label{quotient-narrow-minus}
  0 \to\vermaNminus{h_- - \half,k;\theta}\to
  \verma{h_-,h_-,k;\theta}\to
  \vermaNminus{h_-,k;\theta}\to 0.
  \label{4.9}
\end{equation}
The proof follows from the above observations on the highest-weight
conditions and the character identity
\begin{equation}
  \chi^{\vermaNminus{}}_{h_--\half,k;\theta}(q,z,\zeta) +
  \chi^{\vermaNminus{}}_{h_-,k;\theta}(q,z,\zeta)=
  \chi^{\verma{}}_{h_-,h_-,k;\theta}(q,z,\zeta).
  \label{4.10}
\end{equation}

Similarly, in the Verma module $\verma{h_-,-h_-,k;\theta}$, we have
the charged singular vector $\ket{\Cplus(0, h_-, k;\theta)}$ given by
\begin{equation}
  F^2_{-\theta}\ket{h_-,-h_-,k;\theta}\doteq
  \ketplus{h_--\fhalf,k;\theta}.
  \label{4.11}
\end{equation}
The module generated from this state is
$\vermaNplus{h_--\half,k;\theta}$, and we then have the exact sequence
\begin{equation}%\label{quotient-narrow-plus}
  0\to\vermaNplus{h_- - \half,k;\theta}\to
  \verma{h_-,-h_-,k;\theta}\to
  \vermaNplus{h_-,k;\theta}\to0.
  \label{4.12}
\end{equation}

\section{MFF Singular Vectors in $\hSSL21$ Verma Modules}
The charged singular vectors considered in the previous section are
given by simple explicit formulas because they occur at one of the
extremal states of a Verma module.  In addition to these singular
vectors, there exist those lying in the interior of the module.  We
study these singular vectors in this section.

\subsection{MFF singular vectors as nonvanishing Verma module
  elements} A proposal for $\hSSL21$ singular vectors corresponding to
reflections with respect to two bosonic roots has been known for some
time~\cite{9}~\cite{10}.  The corresponding singular vector formulas
are similar to those for affine Lie algebras given in~\cite{12} by
Malikov, Feigin, and Fuks (hence the acronym MFF), in fact most
similar to the $\hSL2$ singular vector formulas, and we therefore call
them the MFF singular vectors even in the $\hSSL21$ context.  Unlike
their $\hSL2$ counterpart, however, these singular vectors formulas
can give the vanishing result for some ``degenerate'' values of the
highest-weight parameters (this was noted in~\cite{10}).

Because of this vanishing, it thus appears that the \textit{submodule}
otherwise generated from a chosen MFF singular vector ``vanishes'' at
some degenerate points in the highest-weight parameter space.
However, this vanishing in fact pertains to the formula itself and
signals only a possible change of the submodule structure.  The
singular vector formulas in~\cite{9,10} explicitly involve
annihilation operators and can therefore lead to the vanishing result
in the case of certain degenerations, i.e., on certain subsets in the
highest-weight parameter space.  The structure of submodules is
reorganized at these special points of the parameter space, but the
known MFF formula does not provide us with an adequate tool for
studying these degenerations.

In what follows, we propose a formula for the singular vectors (which
we still call the MFF singular vectors) involving only creation
operators and therefore nonvanishing in a given Verma module.
\textit{Generically}, the new expression gives a vector that generates
the same submodule as the ``old'' MFF vector.  Unlike the latter,
however, the new formulas allow us to see how the structure of
submodules is rearranged at the degeneration points in the
highest-weight parameter space.

In fact, each of the ``old'' MFF expressions, $\mffold+$ and
$\mffold-$, is to be replaced with a \textit{pair} of formulas,
respectively $(\mffpp,\mffpm)$ and $(\mffmp,\mffmm)$.  Generically,
the three vectors $\mffpp$, $\mffpm$, and $\mffold+$ generate the same
submodule, but degeneration can affect the relation between the
submodules generated from each of these vectors.  The situation with
the $\mffmp$ and $\mffmm$ vectors is similar, as we describe in what
follows.

\subsection{The MFF singular vectors: ``continued'' expressions and
  the algebraic rules} We recall (see, e.g., \cite{9} and
references therein) that singular vectors occur in a Verma module
$\verma{h_-, h_+,k;\theta}$ whenever $h_-=\jplus(r,s,k)$ or
$h_-=\jminus(r,s,k)$, where
\begin{alignat}{2}
  &\jplus(r,s,k)=\ffrac{r}{2} - \ffrac{s-1}{2}(k+1),&\qquad&r,s\in\oN,
  %\label{jplus}
  \label{5.1}
  \\
  &\jminus(r,s,k)=-\ffrac{r}{2} + \ffrac{s}{2}(k+1),& &r,s\in\oN.
  %\label{jminus}
  \label{5.2}
\end{alignat}
As noted above, we call the corresponding singular vectors the MFF
singular vectors for their obvious similarities with the construction
in~\cite{12}.

To begin with~\eqref{5.1}, we now describe the pair of vectors each
of which generates a submodule in $\verma{\jplus(r,s,k), h_+,k}$.
Writing $\jplus$ for $h_-=\jplus(r,s,k)$ for brevity, we define
\begin{multline}
  \label{5.3}
  \mffpm(r,s,h_+,k) ={}\\
  \begin{aligned}%\label{mff+-}
    {}={}& E^2_{-s}(F^{12}_0)^{2\jplus +
      (2s-2)(k+1)}F^1_{-s+1}\, (E^{12}_{-1})^{2\jplus +
      (2s-3)(k+1)}
    \\
    & \times E^2_{-s+1}(F^{12}_0)^{2\jplus + (2s-4)(k+1)}F^1_{-s+2}\,
    (E^{12}_{-1})^{2\jplus + (2s-5)(k+1)}
    \\
    &
%%\dots\dots\dots\dots\dots\dots\dots\dots\dots\dots\dots\dots\dots\dots
    \qquad\vdots
    \\
    &\times E^2_{-2}(F^{12}_0)^{2\jplus + 2(k+1)}F^1_{-1}\,
    (E^{12}_{-1})^{2\jplus + k+1}
    \\
    &\times E^2_{-1}(F^{12}_0)^{2\jplus}F^1_{0}\,
    \ket{\jplus(r,s,k),h_+,k}
  \end{aligned}
\end{multline}
and
\begin{multline}%\label{mff++}  
  \label{5.4}
  \mffpp(r,s,h_+k)={}\\*
  \begin{aligned}
    {}={}& E^1_{-s}(F^{12}_0)^{2\jplus +
      (2s-2)(k+1)}F^2_{-s+1}\, (E^{12}_{-1})^{2\jplus +
      (2s-3)(k+1)}
    \\
    & \times E^1_{-s+1}(F^{12}_0)^{2\jplus + (2s-4)(k+1)}F^2_{-s+2}\,
    (E^{12}_{-1})^{2\jplus + (2s-5)(k+1)}
    \\
    &
%\dots\dots\dots\dots\dots\dots\dots\dots\dots\dots\dots\dots\dots\dots
    \qquad\vdots
    \\
    &\times E^1_{-2}(F^{12}_0)^{2\jplus + 2(k+1)}F^2_{-1}\,
    (E^{12}_{-1})^{2\jplus + k+1}
    \\
    &\times E^1_{-1}(F^{12}_0)^{2\jplus}F^2_{0}\,
    \ket{\jplus(r,s,k),h_+,k}.
  \end{aligned}
\end{multline}
These expressions can be rewritten as Verma module elements by
repeatedly applying the relations
\begin{gather}%\label{commutef}
  \begin{split}
    (F^{12}_0)^nE^{12}_m &= \bigl(-n (n - 1) F^{12}_m - 2 n
    \Hminus_mF^{12}_0 + E^{12}_m
    F^{12}_0F^{12}_0\bigr)(F^{12}_0)^{n-2},
    \\
    (F^{12}_0)^nE^1_m &= \bigl(-n F^2_m + E^1_mF^{12}_0\bigr)
    (F^{12}_0)^{n-1},
    \\
    (F^{12}_0)^{n}E^2_m &= \bigl(n F^1_m + E^2_mF^{12}_0\bigr)
    (F^{12}_0)^{n-1},
    \\
    (F^{12}_0)^{n}\Hminus_m &= \bigl(n F^{12}_m +
    \Hminus_mF^{12}_0\bigr) (F^{12}_0)^{n-1}
  \end{split}
  \label{5.5}
\end{gather}
and
\begin{equation}%\label{commutee}
  \begin{split}
    (E^{12}_{-1})^nF^{12}_m &= \bigl(-n (n - 1) E^{12}_{m-2} -
    kn\delta_{m,1} E^{12}_{-1} + 2 n \Hminus_{m-1}E^{12}_{-1}\\
    &\kern140pt{} + F^{12}_m E^{12}_{-1}E^{12}_{-1}\bigr)
    (E^{12}_{-1})^{n-2},
    \\
    (E^{12}_{-1})^nF^1_m &= \bigl(n E^2_{m-1} +
    F^1_mE^{12}_{-1}\bigr)(E^{12}_{-1})^{n-1},
    \\
    (E^{12}_{-1})^{n}F^2_m &= \bigl(-n E^1_{m-1} +
    F^2_mE^{12}_{-1}\bigr)(E^{12}_{-1})^{n-1},
    \\
    (E^{12}_{-1})^{n}\Hminus_m &= \bigl(-n E^{12}_{m-1} +
    \Hminus_mE^{12}_{-1}\bigr)(E^{12}_{-1})^{n-1},
  \end{split}
  \label{5.6}
\end{equation}
which are valid for $n\in\oN$ and are postulated for $n\in\oC$.
Useful consequences of these formulas are given by
\begin{multline}%\label{travel-through}
  E^1_{-s}\dots E^1_{-1}\,F^2_{-s}\dots
  F^2_{-1}(E^{12}_{-1})^{\alpha}={}\\
  {}=(-1)^{s+1}E^1_{-s}F^2_{-s}\,(E^{12}_{-1})^{\alpha} E^1_{-s+1}\dots
  E^1_{-1}\,F^2_{-s+1}\dots F^2_{-1} \label{5.7}
\end{multline}
and
\begin{gather}
  E^1_{-s}\dots E^1_{-1}\,F^2_{-s}\dots
  F^2_{-1}\,(F^{12}_{0})^{\alpha} 
  =(F^{12}_{0})^{\alpha}\,
  E^1_{-s}\dots E^1_{-1}\,F^2_{-s}\dots F^2_{-1}.
  %\label{travel-through2}
  \label{5.8}
\end{gather}
and similar relations are valid for the other $\hSL2$ doublet of
currents, $E^2$ and $F^1$.  In what follows, these formulas are
applied for complex~$\alpha$.

By a direct calculation using~\eqref{5.5} and~\eqref{5.6}, it is
easy to verify the following Lemma for the vectors $\mffpp$ and
$\mffpm$.
\begin{Lemma}\label{lemma:mffplus-hw}
  For $r,s\in\oN$, the states $\mffpm(r,s,h_+,k)$ and
  $\mffpp(r,s,h_+,k)$ satisfy the annihilation conditions and
  eigenvalue formulas such that we can write \rom(see~\eqref{3.5}\rom)
  \begin{align}%\label{mff+-hw}
    &\mffpm(r,s,h_+,k)\doteq\ket{\jplus(r,s,k)-r,h_+-s(k+1),k;-s},
    \label{5.9}
    \\
    &\mffpp(r,s,h_+,k)\doteq\ket{\jplus(r,s,k)-r,h_++s(k+1),k;s}.
    %\label{mff++hw}
    \label{5.10}
  \end{align}
\end{Lemma}

These conditions imply, in particular, that these vectors are indeed
``singular'' in the sense that they \textit{generate $\hSSL21$
  submodules}.

Next, in the case described in~\eqref{5.2}, whenever
$h_-=\jminus(r,s,k)\equiv\jminus$, we define the pair of singular
vectors in $\verma{\jminus(r,s,k), h_+,k}$,
\begin{multline}
  \label{5.11}
  \mffmm(r,s,h_+,k)={}\\
  \begin{aligned}%\label{mff--}
    {}={}&(E^{12}_{-1})^{(2s-1)(k+1)-2\jminus}
    \,E^2_{1-s}(F^{12}_{0})^{(2s-2)(k+1)-2\jminus}F^1_{2-s}
    \\
    &\times(E^{12}_{-1})^{(2s-3)(k+1)-2\jminus}
    \,E^2_{2-s}(F^{12}_{0})^{(2s-4)(k+1)-2\jminus}F^1_{3-s}
    \\  
%&\dots\dots\dots\dots\dots\dots\dots\dots\dots\dots\dots\dots\dots\dots
    &\qquad\vdots
    \\
    &\times(E^{12}_{-1})^{3(k+1)-2\jminus}
    \,E^2_{-1}(F^{12}_{0})^{2(k+1)-2\jminus}F^1_{0}
    \\
    &\times (E^{12}_{-1})^{k+1-2\jminus} \ket{\jminus,h_+,k}
  \end{aligned}
\end{multline}
and
\begin{multline}  
  \label{5.12}
  \mffmp(r,s,h_+,k)={}\\
  \begin{aligned}%\label{mff-+}
    {}={}& (E^{12}_{-1})^{(2s-1)(k+1)-2\jminus}
    \,E^1_{1-s}(F^{12}_{0})^{(2s-2)(k+1)-2\jminus}F^2_{2-s}
    \\
    &\times (E^{12}_{-1})^{(2s-3)(k+1)-2\jminus}
    \,E^1_{2-s}(F^{12}_{0})^{(2s-4)(k+1)-2\jminus}F^2_{3-s}
    \\
%&\dots\dots\dots\dots\dots\dots\dots\dots\dots\dots\dots\dots\dots\dots
    &\qquad\vdots
    \\
    &\times (E^{12}_{-1})^{3(k+1)-2\jminus}
    \,E^1_{-1}(F^{12}_{0})^{2(k+1)-2\jminus}F^2_{0}
    \\
    &\times (E^{12}_{-1})^{k+1-2\jminus} \ket{\jminus,h_+,k}.  
  \end{aligned}
\end{multline}
Similarly to Lemma~\ref{lemma:mffplus-hw}, it is easy to obtain the
annihilation conditions satisfied by these vectors.

\begin{Lemma}\label{lemma:mffminus-hw} For $r,s\in\oN$, the states
  $\mffmm(r,s,h_+,k)$ and $\mffmp(r,s,h_+,k)$ satisfy the annihilation
  conditions and eigenvalue formulas such that we can write
  \begin{align}
    &\mffmm(r,s,h_+,k)\doteq
    \ket{\jminus(r,s,k)+r,h_+-(s-1)(k+1),k;1-s}, \label{5.13}
    \\
    &\mffmp(r,s,h_+,k)\doteq
    \ket{\jminus(r,s,k)+r,h_++(s-1)(k+1),k;s-1}.  \label{5.14}
  \end{align}
\end{Lemma}

\subsection{MFF singular vectors: relation between submodules}
We first consider the $\mffpPM$ vectors.  For generic values of $h_+$
and $k$, the $\mffpm(r,s,h_+,k)$ and $\mffpp(r,s,h_+,k)$ singular
vectors generate the same submodule, but this is not so in several
``degenerate'' cases, as we now see.  The relation between $\mffpm$
and~$\mffpp$ is described in the following Lemma.
\begin{Lemma}\label{lemma:mffplus-travel} In the Verma module
  $\verma{\jplus(r,s,k),h_+,k}$, the following relations are satisfied
  up to a sign,
  \begin{multline}%\label{plus-travel-right}
    \label{5.15}
    \underbrace{E^1_{-s}\dots E^1_{s-1}}_{2s}
    \underbrace{\,F^2_{1-s}\dots F^2_{s}}_{2s} \mffpm(r,s,h_+,k)={}
    \\
     =\prod_{i=0}^{2s-1}
    \bigl(h_+-\jplus(r,s,k)-i(k+1)\bigr)\,\mffpp(r,s,h_+,k)    
  \end{multline}
  and
  \begin{multline}%\label{plus-travel-left}
    \label{5.16}
    \underbrace{E^2_{-s}\dots E^2_{s-1}}_{2s}
    \underbrace{F^1_{1-s}\dots F^1_{s}}_{2s} \mffpp(r,s,h_+,k)={}
    \\
     ={\prod_{i=0}^{2s-1}}
    \bigl(h_++\jplus(r,s,k)+i(k+1)\bigr)\,\mffpm(r,s,h_+,k).    
  \end{multline}
\end{Lemma}

This is readily proved using Eqs.~\eqref{5.7}.  These equations imply
that whenever the factor in \textit{either}~\eqref{5.15}
\textit{or}~\eqref{5.16} vanishes, one of the MFF vectors generates a
submodule in the module generated from the other.

A similar relation between $\mffmm$ and~$\mffmp$ is described as
follows.

\begin{Lemma}\label{lemma:mffminus-travel} In the Verma module
  $\verma{\jminus(r,s,k),h_+,k}$, the following relations are
  satisfied up to a sign,
  \begin{multline}%\label{minus-travel-right}
    \label{5.17}
    \underbrace{E^1_{1-s}\dots E^1_{s-2}}_{2s-2}
    \underbrace{F^2_{2-s}\dots F^2_{s-1}}_{2s-2} \mffmm(r,s,h_+,k)={}
    \\    
    ={\prod_{i=1}^{2s-2}}\bigl(h_++\jminus(r,s,k)-i(k+1)\bigr)\,
    \mffmp(r,s,h_+,k)
  \end{multline}
  and
  \begin{multline}%\label{minus-travel-left}
    \label{5.18}
    \underbrace{E^2_{1-s}\dots E^2_{s-2}}_{2s-2}
    \underbrace{F^1_{2-s}\dots F^1_{s-1}}_{2s-2} \mffmp(r,s,h_+,k)={}
    \\    
    ={\prod_{i=1}^{2s-2}}\bigl(h_+-\jminus(r,s,k)+i(k+1)\bigr)\,
    \mffmm(r,s,h_+,k).
  \end{multline}
\end{Lemma}

A number of degenerations that can occur in Verma modules can be
studied by analyzing the relative positions of the singular vectors
$\mffpp$ and$\mffpm$ and, on the other hand, $\mffmp$ and~$\mffmm$.

\subsection{Coexistence of the MFF and charged singular vectors}
\label{sec:coexistence} Explicit formulas for the charged singular
vectors and ``almost'' explicit ones for the MFF singular vectors
allow us to describe the relevant coexistence cases of singular
vectors of these two types.  We start with the Verma module containing
an $\mffold+$ singular vector and a charged singular vector
$\ket{\Cminus(n, h_-, k;\theta)}$ with $n\geq1$; by the choice of the
twisted highest-weight vector from which the module is generated, we
can always assume that $n=1$ (a similar remark applies to other
charged singular vectors considered in what follows).

\begin{Lemma}\label{lemma:coexistence-minus}
  In the Verma module $\verma{h_-, h_- - (k+1), k}$ with
  $h_-=\jplus(r,s,k)$ for $r\geq1$ and $s\geq1$ such that
  $\frac{r}{k+1}\not\in\oZ$, there are submodules
\begin{equation}
    \begin{array}{ccc}
      \verma{h_-, h_- - (k+1),k}&\longleftarrow&\vermaNminus{h_-,k;1}
      \\[2pt]
      \Bigm\uparrow& &\Bigm\uparrow\\
      \verma{h_- - r, h_- - (k+1),k}&\longleftarrow&
      \vermaNplus{h_- - r - \half,k;s},
    \end{array}
    \label{5.19}
  \end{equation}
  where the horizontal arrows are the embeddings of the submodules
  generated from charged singular vectors.
\end{Lemma}

\noindent
{\sc Proof}.  The Verma module $\verma{h_-, h_- - (k+1), k}$ has the
charged singular vector $\Cminus(1,h_-,k)$, with the corresponding
submodule~$\vermaNminus{h_-,k;1}$.  Under the conditions of the Lemma,
$\Cminus(1,h_-,k)$ is the only charged singular vector in $\verma{h_-,
  h_- - (k+1), k}$.  Next, there is a submodule generated from the MFF
vectors $\mffpPM(r,s,h_+,k)$.  It follows from Eqs.~\eqref{5.16}
and~\eqref{5.15} that $\mffpm(r,s,h_+,k)$ and $\mffpp(r,s,h_+,k)$
generate the same submodule. This submodule, in turn, has a charged
singular vector that can be written~as
\begin{multline}
  \cCplus(0,-\ffrac{r}{2}-\ffrac{s-1}{2}(k+1), k;s)\,
  \mffpp(r,s,\ffrac{r}{2}-\ffrac{s+1}{2}(k+1), k)\doteq
  \\
  {}\doteq \ketplus{-\ffrac{r+1}{2}-\ffrac{s-1}{2}(k+1),
    k; s} {}\equiv\ketplus{\jplus(r,s,k) - r - \fhalf, k; s}
\end{multline}
(we remind the reader that the $\doteq$ sign is a statement about the
annihilation and eigenvalue conditions satisfied by this state).
Explicitly, we have (writing $\jplus$ for $\jplus(r,s,k)$ for brevity)
\begin{multline}%\label{in-sub-1}
  \cCplus(0,-\ffrac{r}{2}-\ffrac{s-1}{2}(k+1),k;s)\,
  \mffpp(r,s,\ffrac{r}{2}-\ffrac{s+1}{2}(k+1),k)={}
  \\
   {}=F^2_{-s}\cdot E^1_{-s}(F^{12}_0)^{2\jplus +
    (2s-2)(k+1)}F^2_{-s+1}\, (E^{12}_{-1})^{2\jplus +
    (2s-3)(k+1)}\times
  \\
  \times E^1_{-s+1}(F^{12}_0)^{2\jplus +
    (2s-4)(k+1)}F^2_{-s+2}\, (E^{12}_{-1})^{2\jplus +
    (2s-5)(k+1)}\times
  \\
  \dots\dots\dots\dots\dots\dots\dots
  \dots\dots\dots\dots\dots\dots\dots
  \\
  \times E^1_{-2}(F^{12}_0)^{2\jplus + 2(k+1)}F^2_{-1}\,
  (E^{12}_{-1})^{2\jplus + k+1} E^1_{-1}(F^{12}_0)^{2\jplus}F^2_{0}\,
  \ket{\jplus, \jplus - (k+1), k}.  \label{5.20}
\end{multline}
We now have $F^2_{-s}\,E^1_{-s}(F^{12}_0)^{2\jplus +
  (2s-2)(k+1)}=F^2_{-s}(F^{12}_0)^{2\jplus + (2s-2)(k+1)}E^1_{-s}$ in
accordance with the second formula in~\eqref{5.5}.  This brings the
operators $E^1_{-s}$ and $F^2_{-s+1}$ together in~\eqref{5.20}, and
we can then use the commutation property
\begin{equation*}
E^1_{-s}\,F^2_{-s+1}(E^{12}_{-1})^{2\jplus +
  (2s-3)(k+1)}=E^1_{-s}(E^{12}_{-1})^{2\jplus +
  (2s-3)(k+1)}F^2_{-s+1}.
\end{equation*}
This, in its turn, brings the operators $F^2_{-s+1}$ and
$E^1_{-s+1}$ together, and the process is continued by induction,
until we arrive at $F^2_{-1}\,E^1_{-1}(F^{12}_0)^{2\jplus}F^2_{0}=
F^2_{-1}(F^{12}_0)^{2\jplus}E^1_{-1}F^2_{0}$, which means that the
state~\eqref{5.20} is in the submodule $\vermaNminus{h_-,k;1}$
generated from the charged singular vector $\Cminus(1,h_-,k)=
E^1_{-1}F^2_0\ket{h_-,h_- - (k+1),k}$.  This shows $\vermaNplus{h_- -
  r - \half,k;s}\to\vermaNminus{h_-,k;1}$.\smallskip

Similarly to~\eqref{4.9} and~\eqref{4.12}, we have (as before, with
$h_-=\jplus(r,s,k)$)
\begin{equation}%\label{the-quotient-minus}
  \begin{array}{ccccccccc}
    0&\longleftarrow&\vermaNminus{h_--\half,k;1}
    &\longleftarrow&
    \verma{h_-, h_- - (k+1),k}&\longleftarrow&\vermaNminus{h_-,k;1}
    &\longleftarrow&0
    \\[2pt]
    & &\Bigm\uparrow& &\Bigm\uparrow& &\Bigm\uparrow
    \\
    0&\longleftarrow&\vermaNplus{h_- - r,k;s}
    &\longleftarrow&
    \verma{h_- - r, h_- - (k+1),k}&\longleftarrow&
    \vermaNplus{h_- - r - \half,k;s}
    &\longleftarrow&0.
  \end{array}
  \label{5.21}
\end{equation}
In the quotient with respect to $\vermaNminus{h_-,k;1}$, the submodule
generated from the $\mffpp$ vector is therefore
$\vermaNplus{-\frac{r}{2}-\frac{s-1}{2}(k+1),k;s}$.

We thus see that whenever an $\vermaNminus{}$ submodule is generated
from a charged singular vector in $\verma{\jplus(r,s,k),h_+,k}$, an
$\vermaNplus{}$ submodule appears in the next generation of ``MFF
descendants.''  We now describe this latter case separately, i.e.,
describe the simultaneous occurrence of a $\Cplus(0,h_-,k)$ singular
vector and an MFF vector.
\begin{Lemma}\label{lemma:coexistence-plus}
  In the Verma module $\verma{h_-, -h_-, k}$ with $h_-=\jplus(r,s,k)$
  for $r\geq1$ and $s\geq1$ such that $\frac{r}{k+1}\not\in\oZ$, there
  are submodules
  \begin{equation}%\label{submodule-plus}
    \begin{array}{ccc}
      \verma{h_-,-h_-,k}
      &\longleftarrow&\vermaNplus{h_--\half,k;0}\\[2pt]
      \Bigm\uparrow&&\Bigm\uparrow
      \\
      \verma{h_- - r, -h_- - s(k+1),k;-s}
      &\longleftarrow&\vermaNminus{h_- - r, k;1-s}
    \end{array}
    \label{5.22}
  \end{equation}
\end{Lemma}

\noindent
{\sc Proof}.  The structure of submodules is slightly more interesting
in this case.  The right-hand side of~\eqref{5.16} vanishes for the
current values of the parameters, while the factor on the right-hand
side of~\eqref{5.15} is \textit{non}vanishing under the conditions of
the Lemma.  We thus conclude that $\mffpp(r,s,-h_-,k)$ generates a
submodule in the submodule generated from $\mffpm(r,s,-h_-,k)$.
{}From~\eqref{5.9}, we have
\begin{align}
  \mffpm(r,s,-h_-,k) &\doteq\ket{h_- - r, -h_- - s(k+1),k;-s}=
  \\*
  &=\ket{-\ffrac{r}{2} - \ffrac{s-1}{2}(k+1), -\ffrac{r}{2} -
    \ffrac{s+1}{2}(k+1), k;-s},\notag
\end{align}
and the submodule generated from $\mffpm(r,s,-h_-,k)$ can also be
generated from the corresponding charged singular vector
\begin{equation*}
  \cCminus(1;h_- - r,k;-s)
  \mffpm(r,s,-h_-,k)\doteq
  \ketminus{-\ffrac{r}{2}-\ffrac{s-1}{2}(k+1),k;1-s},
\end{equation*}
which gives the submodule~$\vermaNminus{h_- - r, k, 1-s}$.  The
quotient is then generated from
\begin{gather*}
  F^2_s\mffpm(r,s,-h_-,k)\doteq
  \ketminus{-\ffrac{r+1}{2}-\ffrac{s-1}{2}(k+1),k;1-s}.
\end{gather*}
At the same time, obviously, the module $\verma{h_-, -h_-, k}$ has the
submodule $\vermaNplus{h_- - \half,k;0}$ generated from the charged
singular vector $\Cplus(0,h_-,k)$.  We thus arrive
at~\eqref{5.22}.\smallskip

A key observation for the construction of the resolution is that the
$\verma{}$ submodule in the bottom line in~\eqref{5.22} satisfies the
conditions of Lemma~\ref{lemma:coexistence-minus}, and vice versa, the
$\verma{}$ submodule in the bottom line in~\eqref{5.21} satisfies the
conditions of Lemma~\ref{lemma:coexistence-plus}.

\begin{Rem}\rm
  We note a subtlety in properly identifying submodules
  in~\eqref{5.22}.  We used both the $\mffpm$ and $\mffpp$ vectors,
  with the first one generating the entire submodule $\verma{h_- - r,
    -h_- - s(k+1)}= \verma{-\frac{r}{2}-\frac{s-1}{2}(k+1),
    -\frac{r}{2}-\frac{s+1}{2}(k+1),k;-s}$ and the second generating
  only $\vermaNminus{h_- -
    r}=\vermaNminus{-\frac{r}{2}-\frac{s-1}{2}(k+1), k;1-s}$ (as
  before, we assume $h_- = \jplus(r,s,k)$).  Recalling that an MFF
  singular vector satisfies twisted highest-weight
  conditions~\eqref{5.10}, we thus see that the submodule generated
  from $\mffpp(r,s,-h_-, k)$ is not freely generated in
  $\verma{h_-,-h_-,k}$.  On the other hand, the module freely
  generated from the twisted highest-weight state satisfying the same
  highest-weight conditions as $\mffpp(r,s,-h_-,k)\;$ is
  $\verma{-\frac{r}{2}-\frac{s-1}{2}(k+1), -\frac{r}{2} +
    \frac{3s-1}{2}(k+1),k;s}$, and it has the charged singular vector
  $\Cminus(1-2s,-\frac{r}{2}-\frac{s-1}{2}(k+1),k;s)$, with the
  corresponding submodule $\vermaNminus{-\frac{r+1}{2} -
    \frac{s-1}{2}(k+1),k;1-s}$.  Thus, comparing the module
  $\verma{-\frac{r}{2}-\frac{s-1}{2}(k+1), -\frac{r}{2} +
    \frac{3s-1}{2}(k+1),k;s}$ with the actual submodule
  $\verma{-\frac{r}{2}-\frac{s-1}{2}(k+1),
    -\frac{r}{2}-\frac{s+1}{2}(k+1),k;-s}$ occurring in~\eqref{5.22},
  we see that these are non-isomorphic modules differing by
  transposing the submodule and the quotient (we note in passing that
  the two modules have the same character).  In other words, while
  taking the quotients in~\eqref{5.22} gives
  \begin{equation}%\label{the-quotient-plus}
    \addtolength{\arraycolsep}{-3pt}
    \begin{array}{ccccccccc}
      0\!\!&\longleftarrow&
      \vermaNplus{h_-,k;0}
      &\longleftarrow&\verma{h_-, -h_-,k}
      &\longleftarrow&\vermaNplus{h_- - \half,k;0}
      &\longleftarrow&0\\
      &&\Bigm\uparrow&
      &\Bigm\uparrow&
      &\Bigm\uparrow&&\\
      0&\longleftarrow&
      \vermaNminus{h_- - r - \half, k;1-s}
      &\longleftarrow&
      \verma{h_- - r, -h_- - s(k+1),k;-s}
      &\longleftarrow&
      \vermaNminus{h_- - r, k;1-s}
      &\longleftarrow&0,
    \end{array}
    \label{5.23}
  \end{equation}
  there is a different short exact sequence
  \begin{multline*}
    0\leftarrow
    \vermaNminus{-\frac{r}{2}-\frac{s-1}{2}(k+1), k;1-s}
    \leftarrow
    \verma{-\frac{r}{2}-\frac{s-1}{2}(k+1), -\frac{r}{2} +
      \frac{3s-1}{2}(k+1),k;s}
    \leftarrow\\*
    \leftarrow
    \vermaNminus{-\frac{r+1}{2}-\frac{s-1}{2}(k+1), k;1-s}
    \leftarrow0
  \end{multline*}
  In particular, the throughout mapping 
  \begin{equation*}
  \verma{\jplus(r,s,k), -\jplus(r,s,k),k}\leftarrow
  \verma{-\frac{r}{2}-\frac{s-1}{2}(k+1), -\frac{r}{2} +
    \frac{3s-1}{2}(k+1),k;s}
  \end{equation*}
  \textit{is not an embedding} (i.e.,\ has a kernel).  As can be
  easily verified, there is the isomorphism
  \begin{equation*}
  \verma{-\frac{r}{2}-\frac{s-1}{2}(k+1), -\frac{r}{2} +
    \frac{3s-1}{2}(k+1),k;s}\approx
  \verma{\jplus(r,s,k)-r,-\jplus(r,s,k),k},
  \end{equation*}
  and this untwisted Verma module is therefore \textit{not
    embedded} into $\verma{\jplus(r,s,k), -\jplus(r,s,k),k}$.
\end{Rem}

\begin{Rem}[{\rm the $r=0$ ``degeneration'' of the MFF vectors}]\rm
  %\label{rem:degeneration}
  The exponents in the MFF formulas, e.g., in
  Eqs.~\eqref{5.11}--\eqref{5.12} (which we now rewrite for the spin
  $\jminus(r',s',k)$ for the future convenience) are given by $i(k+1)
  - 2\jminus(r',s',k)$, where $i$ ranges from $1$ to $2s'-1$ (and we
  count the different factors from right to left).  For $i=s'$, the
  corresponding factor is raised to the power~$r'$.  We refer to this
  operator as the \textit{center} of the corresponding MFF formula.
  For definiteness, let now $s'$ be even; then the MFF formula (for
  example,~\eqref{5.11}) has the following structure around its
  center:
  \begin{align*}
    &\dots E^2_{-1-\frac{s}{2}}(F^{12}_0)^{r+2(k+1)}
    F^1_{-\frac{s}{2}} (E^{12}_{-1})^{r + k+1}
    E^2_{-\frac{s}{2}}(F^{12}_0)^{r} F^1_{1-\frac{s}{2}}\times
    \\
    &\qquad\qquad {}\times (E^{12}_{-1})^{r - (k+1)}
    E^2_{1-\frac{s}{2}}(F^{12}_0)^{r-2(k+1)} F^1_{2-\frac{s}{2}}
    \dots\,.  \label{5.24}
  \end{align*}
  We recall that $r'\geq1$ in all the MFF formulas; however, we now
  formally continue this expression to $r'=0$.  We then see that
  $E^2_{-\frac{s'}{2}}F^1_{1-\frac{s'}{2}}$ commutes with
  $E^{12}_{-1}$, and therefore, the two $(E^{12}_{-1})^{\pm(k+1)}$
  factors cancel each other.  Up to an overall sign, therefore,
  \begin{equation}
    \dots
    F^1_{-\frac{s}{2}}
    (E^{12}_{-1})^{k+1}
    E^2_{-\frac{s}{2}}F^1_{1-\frac{s}{2}}
    (E^{12}_{-1})^{-(k+1)}
    E^2_{1-\frac{s}{2}}
    \dots{} =
    \dots F^1_{-\frac{s}{2}}
    E^2_{-\frac{s}{2}}F^1_{1-\frac{s}{2}}
    E^2_{1-\frac{s}{2}}
    \dots\,.
    \label{5.25}
  \end{equation}
  and the same argument can be applied to other factors in the
  singular vector formula, leading to the cancellation of all powers
  of the bosonic generators.  Thus, evaluating~\eqref{5.11} for
  $r'=0$, we find
  \begin{equation}%\label{mff-degenerate}
    \mffmm(0,s,h_+,k)=
    E^2_{1-s}\dots E^2_{-1}\, F^1_{2-s}\dots F^1_{0}
    \,\ket{\frac{s}{2}(k+1),h_+,k}.
    \label{5.26}
  \end{equation}
  This is an extremal state of the module; depending on the value
  of~$h_+$, however, it can become a charged singular vector.
\end{Rem}

\begin{figure}[tbh]
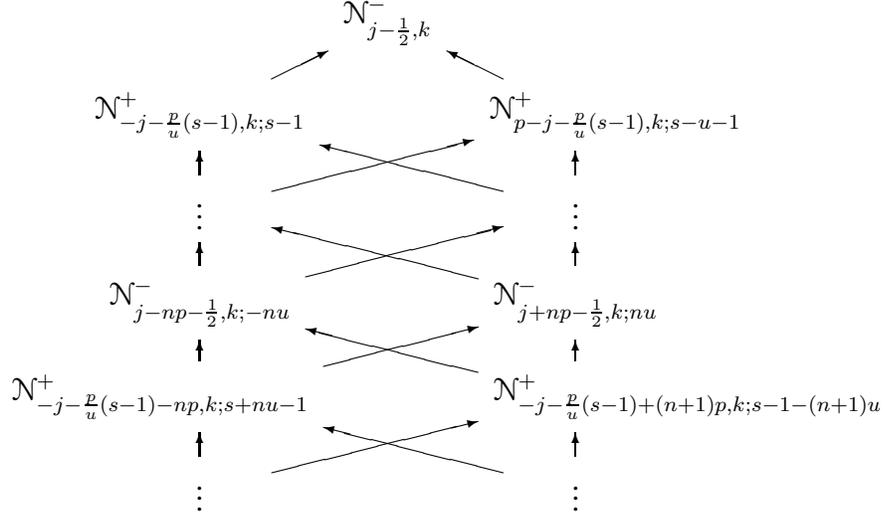

  \begin{equation*}\dgARROWLENGTH=.5em
    \begin{diagram}
      \node[2]{\vermaNminus{j -\half,k}}\\
      \node{\vermaNplus{-j-\frac{p}{u}(s-1), k;s-1}} \arrow{ne}{}
      \node[2]{\vermaNplus{p-j-\frac{p}{u}(s-1),k;s-u-1}
        \kern-30pt}\arrow{nw}\\
      \node{\kern20pt\vdots\kern20pt}\arrow{n}\arrow{ene}
      \node[2]{\kern20pt\vdots\kern20pt}\arrow{n}\arrow{wnw}\\
      \node{\vermaNminus{j - np -\half,k;-nu}} \arrow{n}\arrow{ene}
      \node[2]{\vermaNminus{j+np-\half,k;nu}}
      \arrow{n}\arrow{wnw}\\
      \node{\kern-30pt\vermaNplus{-j-\frac{p}{u}(s-1) - np, k;s + n u
          - 1}} \arrow{n}\arrow{ene}
      \node[2]{\vermaNplus{-j-\frac{p}{u}(s-1)+(n+1)p,k;s-1-(n+1)u}
        \kern-85pt}
      \arrow{n}\arrow{wnw}\\
      \node{\kern20pt\vdots\kern20pt} \arrow{n}\arrow{ene}
      \node[2]{\kern20pt\vdots\kern20pt} \arrow{n}\arrow{wnw}
    \end{diagram}
  \end{equation*}
  \caption[Mappings between narrow Verma modules]{\textsl{Mappings
      between narrow Verma modules.}}\label{fig:grid}
\end{figure}

\section{Conclusions} \label{sec:resolution}
The above statements on mutual positions of singular vectors in
$\hSSL21$ Verma modules give an efficient tool for analyzing the
structure of modules and constructing resolutions of irreducible
representations.  Thus, a repeated application of the lemmas in
Sec.~\ref{sec:coexistence} gives the submodule grid of the module
$\vermaNminus{j -\half,k;1}$ (where $j=\jplus(r,s,k)$, \ 
$k+1={p}/{u}$, and $1\leq r\leq p-1$, \ $1\leq s\leq u$, with coprime
positive integers $p$ and $u$) that entirely consists of narrow Verma
modules, see\ Fig.~\ref{fig:grid}.  This gives the resolution of the
irreducible quotient of the module
\begin{equation*}
\verma{j,j-(k+1),k}=\verma{\frac{r}{2} - \frac{s-1}{2}\frac{p}{u},
  \frac{r}{2} - \frac{s+1}{2}\frac{p}{u}, \frac{p}{u}-1}
\end{equation*} 
through narrow Verma modules.  The construction of the resolution
for the irreducible representation therefore starts with taking the
quotient in diagrams~\eqref{5.21} and~\eqref{5.23}.  The details of
the construction and the resulting character formula will be
considered elsewhere.

\subsection*{Acknowledgments}  We thank B.~Feigin and I.~Tipunin for
interesting discussions.  This work was supported by a Royal Society
grant RCM/ExAgr and partly by the Russian Foundation for Basic
Research Grant No.~01-01-00906 and the Foundation for Promotion of
Russian Science.  AMS~gratefully acknowledges kind hospitality
extended to him at the Department of Mathematical Sciences, University
of Durham.

\end{document}